\documentstyle[11pt,epsf]{article}
\newcommand{\sect}[1]{ \section{#1} \setcounter{equation}{0} }
\newcommand{\half}{\mbox{\small{$\frac{1}{2}$}}} 
\newcommand{\halfsmall}{{\mbox{\footnotesize{$\frac{1}{2}$}}}}  
 
\newcommand{\MSbar}{\overline{\mbox{MS}}} 
\begin{document}
\title{The $\beta$-function of the Wess-Zumino model at $O(1/N^2)$}  
\author{P.M. Ferreira \& J.A. Gracey, \\ Theoretical Physics Division, \\ 
Department of Mathematical Sciences, \\ University of Liverpool, \\ 
Peach Street, \\ Liverpool, \\ L69 7ZF, \\ United Kingdom.}
\date{} 
\maketitle
\vspace{5cm}
\noindent
{\bf Abstract.} We extend the critical point self-consistency method used to 
solve field theories at their $d$-dimensional fixed point in the large $N$ 
expansion to include superfields. As an application we compute the 
$\beta$-function of the Wess-Zumino model with an $O(N)$ symmetry to 
$O(1/N^2)$. This result is then used to study the effect the higher order 
corrections have on the radius of convergence of the $4$-dimensional
$\beta$-function at this order in $1/N$. The critical exponent relating to the 
wave function renormalization of the basic field is also computed to $O(1/N^2)$
and is shown to be the same as that for the corresponding field in the 
supersymmetric $O(N)$ $\sigma$ model in $d$-dimensions. We discuss how the 
non-renormalization theorem prevents the full critical point equivalence 
between both models.  

\vspace{-20cm} 
\hspace{13.5cm} 
{\bf LTH-416} 

\newpage

\sect{Introduction.} 

The Wess-Zumino model is the simplest four dimensional interacting quantum 
field theory which possesses supersymmetry, \cite{wz}. This bose-fermi symmetry 
is believed to play a role in the unification of the forces of nature beyond 
the energy scales accessible by the present generation of accelerators. 
Nevertheless theories with supersymmetry are also of interest for testing out 
and developing new ideas. One reason for this is that one can calculate much 
more easily in the quantum supersymmetric theory compared with say its bosonic 
sector, due to cancellations between graphs involving bosons and fermions. For 
example, there is no two loop term in the $\MSbar$ $\beta$-function of the two 
dimensional supersymmetric $\sigma$ model on an arbitary manifold, \cite{alv}. 
In the corresponding bosonic $\sigma$ model the two loop contribution to the 
$\beta$-function is non-zero, \cite{sig}. Moreover, in certain situations, like
the Wess-Zumino model, the renormalization is constrained in such a way that 
the calculation of the coupling constant renormalization is reduced to 
obtaining the wave function renormalization. In other words ensuring unbroken 
supersymmetry in the quantum theory restricts the form of possible counterterms
in the superpotential and hence determines a renormalization constant, 
\cite{wz,nonren}. This non-renormalization theorem allows one to compute higher
order corrections rather easily. For instance, the four loop $\beta$-function 
for the pure model with one superfield is known, \cite{earlyloop,grisa,avdeev},
as well as that for its extension to include superfields with general internal
symmetries, \cite{wzssm}. Although the Wess-Zumino model was originally 
formulated in terms of ordinary spin-$0$ and $\halfsmall$ fields, there is a 
further calculational tool which substantially reduces the amount of work 
needed for these higher order calculations. This is attained by grouping the 
component fields in the same multiplet into superfields where the Lorentz 
space-time is extended to include Grassmann coordinates. Consequently in this 
superspace the Feynman supergraphs group classes of component field diagrams 
together thereby reducing the number that need to be computed. Moreover 
performing the integrations over the Grassmann variables, known as the 
$D$-algebra, the resulting Feynman integrals are invariably simple scalar 
integrals. This property can be better appreciated in superfield calculations 
in supersymmetric gauge theories where the number of component fields is larger
than in the Wess-Zumino model and high loop integrals involving the component 
gauge field become extremely tedious compared with a scalar integral.  

Recently some insight into the structure of the Wess-Zumino model 
$\beta$-function beyond the present four loop level has been gained in 
\cite{bubble} through using the conventional large $N$ bubble sum method. The 
leading order corrections in $1/N$ were computed for the case where the 
Wess-Zumino model has a multiplet of $O(N)$ fields coupled to a separate scalar
superfield. One of the motivations of that study was to ascertain the large 
coupling constant behaviour of the $\beta$-function in a four dimensional model
and hence to understand some of the problems in its coupling constant 
resummation, as well as being the foundation for repeating the exercise for a 
supersymmetric gauge theory. As leading order calculations do not reveal a 
substantial amount of structure it would be interesting to push the $1/N$ 
calculations to a higher order in both the Wess-Zumino and supersymmetric gauge
theories. However, a formalism does exist to achieve this and is based on the 
critical renormalization group and $d$-dimensional conformal field theory 
methods which were pioneered in the series of papers of ref. \cite{vas1,vas2} 
for the $O(N)$ $\sigma$ model in two dimensions. Before attempting to apply 
such methods to these models there are several problems which need to be 
solved. Although the methods of \cite{vas1,vas2} have already been applied to 
the supersymmetric $O(N)$ $\sigma$ model in \cite{susysig1,susysig2} to obtain 
the $\beta$-function at $O(1/N^2)$ and the supersymmetric $CP(N)$ $\sigma$ 
model at $O(1/N)$, \cite{susycpn}, these calculations were performed using the 
component lagrangians. Indeed the computation of $77$ component Feynman 
diagrams in \cite{susysig2} to deduce the $\beta$-function at $O(1/N^2)$ only 
serves to indicate that the development of a superfield approach would yield a 
more compact approach. This is one of the aims of this paper and will be 
provided for the Wess-Zumino model. As an application we will then compute its 
$\beta$-function at a new order, $O(1/N^2)$, as a function of the space-time 
dimension, $d$. Briefly, the method involves determining the critical exponents
of the theory at the phase transition defined by the non-trivial zero of the 
$d$-dimensional $\beta$-function. As there is no mass at such a point the 
propagators of all the fields, for instance, involve the momentum raised to 
some power which is known as a critical exponent. From a field theoretic point 
of view this is a fundamental object as it is renormalization group invariant 
and therefore physically measurable. It depends only on the space-time 
dimension and the parameters of any internal symmetries. For our purpose this 
will be $N$. Therefore it can be expanded in powers of $1/N$ when $N$ is large 
and deduced order by order by examining the scaling behaviour of the particular
Schwinger Dyson equations truncated at the appropriate order. Moreover, as the 
exponents will depend on $d$ $=$ $4$ $-$ $2\epsilon$ they can be related 
through the critical renormalization group to the corresponding renormalization
constants one ordinarily computes in perturbation theory. Hence one can deduce 
new coefficients in the perturbative series in successive orders in $1/N$ by 
performing the $\epsilon$-expansion of the exponents beyond the order currently
known. 
 
Finally, another motivation for studying the Wess-Zumino model at this
$d$-dimensional critical point is to understand the universality class to which
it belongs. By this we mean the following. The bosonic sector of the $O(N)$ 
Wess-Zumino model is effectively $\phi^4$ theory with an $O(N)$ symmetry. The 
field which is not an element of the multiplet $O(N)$ acts as auxiliary which 
if eliminated yields the usual quartic interaction. As is well known (see, for 
example, \cite{zinn}) in $d$-dimensions $O(N)$ $\phi^4$ theory lies in the same
universality class as the lower dimensional $O(N)$ bosonic $\sigma$ model and 
therefore the critical exponents one computes in either model are equivalent. 
Indeed these are of interest because both models underlie the physics of the 
phase transition of the Heisenberg ferromagnet. Therefore we are motivated to 
determine if there is an analogous lower dimensional equivalent theory for the 
Wess-Zumino model. Intuitively one would expect it to have some connection with
the supersymmetric $O(N)$ $\sigma$ model. To achieve this we need to compare 
the values for the Wess-Zumino exponents with those of 
\cite{susysig0,susysig1,susysig2}. Such an equivalence would therefore play a 
role in some supersymmetric extension of the Heisenberg ferromagnet in three 
dimensions, though of course it is not clear if such an object exists in the 
real physical world.  

The paper is organised as follows. We review the relevant properties of the 
Wess-Zumino model in section 2 and introduce the superfield approach for the
critical point large $N$ method. This is used to reproduce the known results at
leading order in $1/N$ in section 3 where the $O(1/N^2)$ correction is also
derived for the wave function renormalization. The higher order correction to
the $\beta$-function is determined in section 4 where the technical details of
the computation of the three and four loop diagrams which occur are also 
discussed. Finally we collect our results in section 5 and produce several new
coefficients that will appear in the $\MSbar$ $\beta$-function at five and 
six loops. 

\sect{Critical point formalism.}

We will be working with an $O(N)$ symmetric Wess-Zumino model containing one
chiral superfield $\sigma$ and $N$ chiral superfields $\Phi^i$, $1$ $\leq$ $i$ 
$\leq$ $N$. Its action is given by 
\begin{equation}
S ~=~ \int \, d^d x \, \left[ \, \int d^4 \theta \, \left( \bar{\Phi}^i\Phi^i 
+ \frac{\bar{\sigma}\sigma}{g^2} \right) ~-~ \frac{1}{2} \int d^2 \theta \, 
\sigma \Phi^2 ~-~ \frac{1}{2} \int d^2 \bar{\theta} \, \bar{\sigma} 
\bar{\Phi}^2 \right]  
\label{eq:act}  
\end{equation}
The coupling constant $g$ has been rescaled into the $\sigma$-field so that the
$3$-vertex is in the form which easily allows us to use the technique of 
uniqueness, \cite{par,vas2}, to compute the higher order graphs which will 
arise. Moreover the chirality of the fields in (\ref{eq:act}) will prove to be 
useful in substantially reducing the number of possible diagrams which we will
need to consider then. Before carrying out a dimensional analysis of the action
to ascertain the canonical dimensions of the fields and the couplings, which is
the first step in the critical point formalism, we recall some of the 
practicalities of the calculation of the perturbative $\beta$-function. If we
define a new field $\sigma^\prime$ by $\sigma^\prime$ $=$ $\sigma/g$ then one
would recover the conventional form of the interaction term. The 
non-renormalization theorem of \cite{wz,nonren} implies that the vertex is 
finite and so the overall vertex renormalization constant involves no 
infinities. Therefore the renormalization group functions of the rescaled 
interaction and its constituent fields satisfy the following simple relation 
\begin{equation}  
\beta(g) ~=~ [ 2\gamma_\Phi(g) ~+~ \gamma_{\sigma^\prime}(g) ] g 
\label{eq:betdef} 
\end{equation} 
which implies the wave function renormalization purely determines how the 
coupling constant runs. The result (\ref{eq:betdef}) was used in 
\cite{earlyloop,grisa,avdeev} to deduce $\beta(g)$ for the model without any  
internal symmetries. As we are dealing with the model with an $O(N)$ symmetry 
we require the perturbative results as a function of $N$. To achieve this we 
have specialised the four loop result of \cite{wzssm} to the $O(N)$ model. 
Useful in checking the relevant symmetry factors in this case was the package 
{\sc Qgraf}, \cite{qgraph}. Therefore, the $\MSbar$ results are
\begin{eqnarray}
\gamma_{\Phi}(g) &=& g ~-~ \frac{1}{2} (N+2) g^2 ~-~ \frac{1}{4} 
\left( N^2 - 10N - 4 - 24\zeta(3) \right) g^3 \nonumber \\  
&&-~ \left( \frac{1}{24} (3N^3 + 16N^2 + 152N + 40) - \frac{\zeta(3)}{4} 
(N^2 - 4N - 36)(N+4) \right. \nonumber \\ 
&& \left. ~~~~~~ -~ 3\zeta(4)(N+4) + 20\zeta(5)(N+2) \frac{}{} \right) g^4
{} ~+~ O(g^5)  
\label{eq:gammaphi}
\end{eqnarray}
and 
\begin{eqnarray} 
\gamma_{\sigma^\prime}(g) &=& \frac{N}{2} g ~-~ Ng^2 ~+~ \frac{N}{2}( 2N + 1 
+ 6\zeta(3) )g^3 \nonumber \\ 
&& +~ N \left( \frac{1}{12} (5N^2 - 56N - 16) - \frac{1}{2} \zeta(3) 
(N^2+6N+44) \right. \nonumber \\ 
&& \left. ~~~~~~~~ +~ \frac{3}{2} \zeta(4) (N+4) - 10\zeta(5) (N+2) \right) g^4 
{}~+~ O(g^5)  
\label{eq:gammasig}
\end{eqnarray} 
Hence, (\ref{eq:betdef}) implies  
\begin{eqnarray}
\beta(g) &=& \frac{1}{2} (d-4)g ~+~ \frac{1}{2} (N+4) g^2 ~-~ 2 (N+1)g^3 
{}~+\, \left( \frac{1}{2} (N^2+11N+4) + 3(N+4)\zeta(3) \right) g^4 \nonumber \\ 
&& +~ \left( \frac{1}{6} (N^3 - 36N^2 - 84N - 20) - 3(N^2+16N+24)\zeta(3) 
\right. \nonumber \\ 
&& \left. ~~~~~~ +~ \frac{3}{2}(N+4)^2\zeta(4) - 10(N+2)(N+4)\zeta(5) 
\frac{}{} \right) g^5 ~+~ O(g^6) 
\label{eq:beta}
\end{eqnarray}
We note that for various reasons we have suppressed the usual factors of $\pi$ 
associated with each power of the coupling in these expressions since, for 
example, they can readily be restored by a simple rescaling. Another reason is
related to the critical point approach where we deduce critical exponents 
which correspond through the renormalization group to these renormalization 
group functions. The critical exponents are more fundamental quantities than
the renormalization group functions in that they are renormalization group 
invariant. Therefore they take the same values in any mass independent 
renormalization scheme. This criterion arises from the fact that at a fixed 
point of the $\beta$-function there is scaling and conformal invariance which
would be broken by any non-zero mass. Therefore since, for instance, MS and 
$\MSbar$ are two examples of mass independent renormalization schemes the 
critical exponents deduced in either scheme will be the same. The only 
difference is that the value of the critical couplings in each scheme will not
be the same but they will be simply related by a constant rescaling of either 
of the couplings. However, throughout we will always have in mind that our 
results will relate to the $\MSbar$ scheme.   

With these renormalization group functions we can now develop the critical 
point formalism. We take the $d$-dimensional $\beta$-function of 
(\ref{eq:beta}) and compute the location of the non-trivial $d$-dimensional
fixed point as  
\begin{eqnarray} 
g_c &=& \frac{2\epsilon}{N} ~+~ \left( - \, 8\epsilon + 16\epsilon^2 
- 8\epsilon^3 - \frac{16}{3}\epsilon^4 + O(\epsilon^5) \right) \frac{1}{N^2} 
\nonumber \\  
&& +~ \left( \frac{}{} 32\epsilon - 176\epsilon^2 - 8(6\zeta(3) - 37)\epsilon^3
\right. \nonumber \\ 
&& \left. ~~~~~~ + \frac{16}{3}( 60\zeta(5) - 9\zeta(4) + 18\zeta(3) - 4 ) 
\epsilon^4 + O(\epsilon^5) \right) \frac{1}{N^3} ~+~ 
O \left( \frac{\epsilon}{N^4} \right) 
\label{eq:critcpl} 
\end{eqnarray} 
in powers of $1/N$ where $d$ $=$ $4$ $-$ $2\epsilon$. Although supersymmetry is
not easily defined in $d$-dimensions we note that this is a valid step since in
perturbative $\MSbar$ calculations with a dimensional regulator (\ref{eq:beta})
would be produced as the penultimate step in determining the $4$-dimensional 
$\beta$-function. Moreover it was shown in \cite{susysig1,susysig2} that the 
$d$-dimensional version of the supersymmetric $\sigma$ model preserved certain 
features of the symmetry at its corresponding fixed point and this allows us to 
make similar assumptions about the Wess-Zumino model in $d$-dimensions. For 
instance, the component fields of each boson and fermion partner in a 
supermultiplet retained the {\em same} anomalous dimension in arbitrary 
dimensions. With (\ref{eq:critcpl}) we can now dimensionally analyse the 
$d$-dimensional action (\ref{eq:act}) at $g_c$ choosing to work with the 
$\sigma$-field. Unlike the component calculation of \cite{susysig1} we define 
the canonical dimensions of the fields in relation to the full superspace and 
take the dimension of the Grassmann coordinates as $-\,1/2$. As the 
renormalization group functions are non-zero we must also allow for the 
possibility of anomalous contributions which we will define with respect to the
usual conventions of \cite{vas1}. Defining the full dimension of the $\Phi$ 
superfield as $\alpha$ then the quadratic term of (\ref{eq:act}) implies we set 
\begin{equation}
\alpha ~=~ \mu ~-~ 1 ~+~ \frac{\eta}{2} 
\end{equation}
where $\eta$ is the critical exponent corresponding to the wave function 
renormalization at criticality. In other words $\eta$ $=$ $\gamma_\Phi(g_c)/2$.
From the interaction term, if we denote by $\beta$ the full dimension of the 
$\sigma$ superfield and by $\chi$ the vertex dimension, then 
\begin{equation} 
\beta ~=~ 2\mu ~-~ 1 ~-~ 2\alpha ~-~ \chi 
\end{equation} 
However, as the non-renormalization theorem implies the vertex is finite to
all orders in perturbation theory then the (composite) operator $\sigma\Phi^2$
can have no anomalous piece. Therefore we set $\chi$ $=$ $0$. This observation 
will have a simplifying effect in, for example, the computation of the $1/N^2$
corrections to the Schwinger Dyson equations we will solve. Hence 
\begin{equation} 
\beta ~=~ 1 ~-~ \eta 
\end{equation} 
and the vertices of all the Feynman diagrams will in principle be one step 
away from uniqueness. From the remaining term of the action we deduce a scaling
law relating the dimension of the $\sigma^2$ composite operator to the 
$\beta$-function exponent. As $\beta(g_c)$ $=$ $0$ this is defined to be 
$\omega$ $=$ $-$ $\beta^\prime(g_c)$. Therefore, we have 
\begin{equation} 
\omega ~=~ 2\mu ~-~ 2 ~-~ 2\beta ~+~ \eta_{\sigma^2} 
\label{eq:compdef} 
\end{equation} 
where $\eta_{\sigma^2}$ is the anomalous dimension of the bare composite 
operator. Although one can use (\ref{eq:compdef}) to determine the 
$\beta$-function by calculating the anomalous dimension of $\sigma^2$ we have
chosen to use the method of \cite{vas1,vas2} of solving the Dyson equations.
It turns out that for scalar theories at least, both ways are related in that 
the Feynman diagrams where one inserts $\sigma^2$ in a $\sigma$ $2$-point 
function and examines the resulting diagrams, it is easy to see that they have 
a complete analogue with the diagrams which arise in the Schwinger Dyson 
approach. However, both methods require a similar amount of effort to determine
the solution.  

Near $g_c$, we can define the asymptotic behaviour of the dressed propagators 
of $\Phi$ and $\sigma$ in Euclidean space. As $p$ $\rightarrow$ $\infty$, 
\begin{eqnarray}
\langle \bar{\Phi}(-p,\theta) \Phi(p,\theta^\prime) \rangle & \sim & 
\frac{A \delta^4 (\theta - \theta^\prime)}{(p^2)^{\mu - \alpha}} \nonumber \\  
\langle \bar{\sigma}(-p,\theta) \sigma(p,\theta^\prime) \rangle & \sim & 
\frac{B \delta^4 (\theta - \theta^\prime)}{(p^2)^{\mu - \beta}} 
\label{eq:sca}
\end{eqnarray}
where we have set $d$ $=$ $2\mu$ and $A$ and $B$ are unknown momentum 
independent amplitudes. Corrections to these scaling forms can also be defined.
For example, we take 
\begin{eqnarray}
\langle \bar{\Phi}(-p,\theta) \Phi(p,\theta^\prime) \rangle & \sim & 
\frac{A\delta^4 (\theta - \theta^\prime)}{(p^2)^{\mu - \alpha}} 
\left[ 1 ~+~ \frac{A^\prime}{(p^2)^{\omega}} \right] \nonumber \\ 
\langle \bar{\sigma}(-p,\theta) \sigma(p,\theta^\prime) \rangle & \sim & 
\frac{B\delta^4 (\theta - \theta^\prime)}{(p^2)^{\mu - \beta}} 
\left[ 1 ~+~ \frac{B^\prime}{(p^2)^{\omega}} \right]
\label{eq:prop}
\end{eqnarray}
In principle one can include other correction terms involving other exponents
with different canonical dimensions. However, renormalizability and 
hyperscaling laws ensure that knowledge of two independent exponents such as 
$\eta$ and $\nu$ is sufficient to deduce the others. Here $\omega$ is the 
exponent which relates to corrections to scaling. The two pairs of amplitudes 
$(A,B)$ and $(A^\prime,B^\prime)$ are independent of one another. Following the
methods of \cite{vas1,vas2}, we will use (\ref{eq:prop}) to solve the Dyson 
equations. The inclusion of a non-zero anomalous dimension in the exponents of 
the propagators means one considers only those diagrams with no dressing on the
propagators which of course will further reduce the number of graphs which 
would need to be computed at high orders. 

Ordinarily in the critical point method of \cite{vas1,vas2} the next stage is
to determine the asymptotic scaling forms of the inverses to (\ref{eq:sca}) and
(\ref{eq:prop}) by inverting them in momentum space. This is because they will 
appear in the Dyson equation as illustrated in fig. 1. However, the situation 
is different here and can be best introduced by first recalling the nature of 
perturbative calculations in superspace. In the calculation of the corrections 
to, say, the quadratic terms in the effective action one needs to retain the 
external momentum space superfields explicitly when representing the graph from
the super-Feynman rules. The reason for this is that the interactions and 
propagators involve supercovariant derivatives which act on the Grassmann 
coordinates and hence the external fields since they too depend on $\theta$. In 
ordinary non-supersymmetric calculations it is not necessary to include the
external fields since it is the case that for (local) quantum field theories 
derivative couplings are defined in coordinate space. Therefore when performing
computations in momentum space their Fourier transform corresponds to vectors 
of the conjugate variable which therefore do not act differentially on the 
external legs. After the manipulation of the supercovariant derivatives, known 
as performing the $D$-algebra, one is left with a simple momentum space Feynman
integral which can in principle be computed. The final result is an integral 
over the full momentum superspace which includes the fields corresponding to 
the external legs. For a $2$-point function this would be of the form, 
\cite{super,grisa},  
\begin{equation}  
\int d^d p \int d^4 \theta \, f(p) \bar{\Phi}(-p,\theta) \Phi(p,\theta) 
\label{eq:formal} 
\end{equation} 
where $\Phi$ and $\bar{\Phi}$ are the external fields whose momentum is $p$. 
Therefore this would be regarded as the correction to the corresponding term in
the effective action. From the point of view of perturbative renormalization 
the extraction of the pole part in $f(p)$ with respect to the regularization 
would then be absorbed by the appropriate counterterm. In light of this the 
critical point approach to the Dyson equations is adapted to be similar. One 
manipulates the graphs in momentum superspace using ordinary $D$-algebra but 
with the lines of the graphs of fig. 1 replaced by the critical propagators, 
(\ref{eq:sca}) and (\ref{eq:prop}), and the external fields assumed to be 
included. One again obtains an expression of the form (\ref{eq:formal}) but the
corresponding $f(p)$ will now depend on the exponents of the lines, $\alpha$ 
and $\beta$, of the original graph. The momentum dependence will be in scaling 
form with the exponent given by the dimension of the original integral. The 
remaining part of (\ref{eq:formal}) would be regarded as the propagator. In the
approach of \cite{vas1,vas2} in non-supersymmetric calculations this would not 
appear {\em here} since the fields of the external legs are not included which 
is the reason why one needs to invert the corresponding (\ref{eq:sca}). So in 
the superspace approach this is not necessary since the propagator appears 
naturally with the scaling part of the graph. Although we have included the 
inverses in fig. 1, these are also meant to account for the situation when one 
has the corrections to scaling of (\ref{eq:prop}). Therefore, we define 
\begin{equation}
\Phi^{-1}(p) ~ \sim ~ \frac{1}{A} \left[ 1 ~-~ \frac{A^\prime}{(p^2)^{\omega}} 
\right] ~~~,~~~ \sigma^{-1}(p) ~ \sim ~ \frac{1}{B} \left[ 1 ~-~ 
\frac{B^\prime}{(p^2)^{\omega}} \right]
\label{eq:sigphi}
\end{equation}
which is similar to the situation in \cite{vas1,vas2}. In practical 
calculations it will be a simple exercise to verify that the above argument 
leads to dimensionally consistent equations.  

\sect{Solving the Dyson equations.}

We are now in a position to determine expressions for the critical exponents
and begin with the leading order analysis. With the propagators from 
(\ref{eq:prop}), the Dyson equations, which are given in fig. 1 and have been
truncated at leading order in $1/N$, may be written in the following integral 
form in the scaling region as
\begin{eqnarray}
0 &=& \Phi^{-1}(p) ~+~ AB {\cal H}(\Phi,p) \int d^{2\mu}k \, 
\frac{1}{(k^2)^{\mu - \alpha} ((k - p)^2)^{\mu - \beta}} \left[ 1 ~+~ 
\frac{A^\prime}{(k^2)^{\omega}} ~+~ \frac{B^\prime} {((k - p)^2)^{\omega}} 
\right] \nonumber \\  
0 &=& \sigma^{-1}(p) ~+~ \frac{1}{2} N A^2 {\cal H}(\sigma,p) \int d^{2\mu}k \,
\frac{1}{(k^2)^{\mu - \alpha} ((k - p)^2)^{\mu - \alpha}} 
\left[ 1 ~+~ \frac{A^\prime}{(k^2)^{\omega}} ~+~ \frac{A^\prime} 
{((k - p)^2)^{\omega}} \right] \nonumber \\  
\label{eq:sdeqn} 
\end{eqnarray}
where the functions ${\cal H}$ of the fields and external momentum represent 
the $\theta$ integral of (\ref{eq:formal}) and we have omitted terms which are 
quadratic in $A^\prime$ and $B^\prime$. We note that we use a form of the Dyson
equations where the propagators are dressed but not the vertices. The factor of
$\halfsmall$ comes from the symmetry of the first graph in fig. 1. Substituting
for $\alpha$, $\beta$ and $\omega$ and decoupling (\ref{eq:sdeqn}) into terms 
with the same momentum dimension, we obtain  
\begin{eqnarray}
0 &=& 1 ~+~ z \nu \left( \mu - \half \eta, \mu - 1 + \eta, 1 - \half \eta 
\right) \nonumber \\ 
0 &=& 1 ~+~ \half Nz\nu \left( 2\mu - 2 + \eta, 1 - \half \eta, 
1 - \half \eta \right) 
\label{eq:eta}
\end{eqnarray}
\begin{eqnarray}
0 &=& \left[ \nu  \left( \mu - 1 + \omega^\prime - \half \eta, \mu - 1 + \eta, 
2 - \half \eta - \omega^\prime \right) \,z - 1 \right]\,A^\prime \nonumber \\
&& ~+~ \nu \left( 2\mu - 3 + \omega^\prime + \eta, 2 - \omega^\prime - \half 
\eta, 1 - \half \eta \right) zB^\prime \nonumber \\
0 &=& zN \nu \left( \mu - 1 + \omega^\prime - \half \eta, 1 - \half \eta, 
\mu + \eta - \omega^\prime \right) A^\prime ~-~ B^\prime 
\label{eq:om}
\end{eqnarray}
where we have set $z$ $=$ $A^2B$ and defined the anomalous piece of $\omega$ by
$\omega$ $=$ $\mu$ $-$ $2$ $+$ $\omega^\prime$. The function $\nu$ is defined 
for arbitrary   $\alpha$, $\beta$ and $\gamma$ as $\nu(\alpha,\beta,\gamma)$ 
$=$ $a(\alpha) a(\beta) a(\gamma)$ with $a(\alpha)$ $=$ 
$\Gamma(\mu - \alpha)/\Gamma(\alpha)$. Eliminating $z$ from (\ref{eq:eta}) we 
obtain the relation that will give us $\eta$ at leading order in $1/N$ 
\begin{equation}
\eta ~=~ \frac{4\Gamma\left(1+\eta/2\right)}{N\Gamma\left(
\mu-\eta/2\right)} \nu \left(2-\mu-\eta,\mu-1+\half\eta, \mu-1+\eta \right) 
\label{eq:de}
\end{equation}
Setting $\eta$ $=$ $\sum_{i=1}^{\infty} \eta_i/N^i$ and expanding (\ref{eq:de})
in powers of $1/N$ we find $\eta_1$ is 
\begin{equation}
\eta_1 ~=~ \frac{4\Gamma(2\mu-2)}{\Gamma(\mu) \Gamma^2(\mu-1) \Gamma(2-\mu)}
\label{eq:eta1}
\end{equation} 
Remarkably the computation of $\eta_2$ requires only the expansion of 
(\ref{eq:de}) to the next order in $1/N$. Ordinarily one would have two and 
three loop corrections to each of the Dyson equations of fig. 1 but in the 
Wess-Zumino model the chirality condition on the vertex excludes these. 
Therefore the first non-zero corrections will arise at $O(1/N^3)$. In this 
respect the Hartree-Fock approximation at $O(1/N^2)$ determines the full value 
of $\eta_2$. Therefore we simply obtain
\begin{equation}
\eta_2 ~=~ \eta_1^2 \left[ \psi(2-\mu) ~+~ \psi(2\mu-2) ~-~ \psi(\mu-1) ~-~ 
\psi(1) ~+~ \frac{1}{2(\mu-1)}\right]
\label{eq:eta2}
\end{equation}
where $\psi(x)$ is the logarithmic derivative of the Euler $\Gamma$-function. 
Moreover the absence of these higher order contributions means that unlike the 
bosonic $\sigma$ model calculation of \cite{vas1,vas2} the Dyson equation is
finite to this order and does not need to be either regularized or 
renormalized. Regularization is usually necessary due to vertex subgraph
divergences in the higher order graphs as the vertices of the Dyson equation 
are undressed. It is introduced by shifting the analogous $\sigma$-field 
exponent by an infinitesimal amount, $\Delta$, which is equivalent to setting 
$\chi$ $\rightarrow$ $\chi$ $+$ $\Delta$. As there is no vertex renormalization
in the Wess-Zumino model and since the critical large $N$ method is effectively
a perturbative expansion in the vertex anomalous dimension then we are led to
reason that at the fixed point the Dyson equations do not require any 
renormalization at {\em any} order in $1/N$. For subsequent calculations we 
will require the value of $z$ to two orders. Also setting $z$ $=$ 
$\sum_{i=1}^\infty z_i/N^i$ knowledge of $\eta_1$ and $\eta_2$ allows us to 
deduce 
\begin{eqnarray} 
z_1 &=& -~ \half \Gamma(\mu) \eta_1 \nonumber \\ 
z_2 &=& \half \Gamma(\mu) \left[ \psi(2-\mu) ~+~ \psi(2\mu-2) ~-~ 
\psi(\mu-1) ~-~ \psi(1) \right] \eta_1^2 
\label{eq:zdefn} 
\end{eqnarray}  

The second pair of equations, (\ref{eq:om}), determines corrections to the 
exponent $\omega$. If one formally writes (\ref{eq:om}) in matrix form 
\begin{equation}
\left( 
\begin{array}{cc}
x & y \\
v & w 
\end{array}
\right)  
\left( 
\begin{array}{c}
A^\prime \\ B^\prime 
\end{array}
\right) ~=~ 0
\label{eq:omdet} 
\end{equation}
then $\omega$ is determined by requiring that the determinant of this matrix
vanishes, \cite{vas2}. This gives
\begin{eqnarray}
1 &=& z^2N \nu \left( 2 \mu- 3 + \omega^\prime + \eta, 1 - \half\eta, 2 - 
\omega^\prime - \half \eta \right) \nu \left( \mu - 1 + \omega^\prime 
- \half \eta, 1 - \half \eta, \mu + \eta - \omega^\prime \right) \nonumber \\
&& +~ z \nu \left( \mu - 1 + \omega^\prime - \half \eta, \mu - 1 + \eta, 
2 - \half \eta - \omega^\prime \right) 
\label{eq:omt}
\end{eqnarray}
However, in deducing (\ref{eq:omt}) one needs to bear in mind that the extra
graphs at $O(1/N^2)$, which would usually play a role in the evaluation of 
$\eta_2$ but which are absent due to chirality, would also in principle need to
be included here. Though they would only enter in one of the Dyson equations
because the leading $N$-dependence of each of the entries of the matrix of 
(\ref{eq:omdet}), when considered in coordinate space, is not the same. 
Moreover this feature has already been observed and discussed in similar 
calculations in other models, \cite{gn2}. Although the relevant graphs are 
absent in determining the first non-trivial term of $\omega$ certain topologies
which cannot be ruled out by chirality occur at higher order and therefore will
need to be included in the relevant Dyson equation of (\ref{eq:omdet}). 
Therefore setting $\omega^\prime$ $=$ $\sum_{i=1}^{\infty} \omega_i/N^i$ and 
retaining only the leading terms in (\ref{eq:omt}), we obtain
\begin{equation}
1 ~=~ -~ \frac{(2\mu-3)\eta_1}{(\omega_1-\eta_1)} ~-~ \frac{(\mu-1)\eta_1}
{2(\mu-2)N} 
\label{eq:om1} 
\end{equation}
As the second term is suppressed by a factor of $1/N$ compared with the first 
we can ignore it and rearrange (\ref{eq:om1}) to find 
\begin{equation}
\omega_1 ~=~ -~ 2(\mu-2)\eta_1
\end{equation}

We need to compare these results for $\eta_1$ and $\omega_1$ with the results
already obtained via the conventional bubble sum methods, \cite{bubble}, and 
demonstrate their equivalence. To do so we must convert the critical exponents 
into the original renormalization group functions. From the definition of the 
critical coupling, (\ref{eq:critcpl}), we can replace the $\mu$-dependence in 
(\ref{eq:eta1}) at leading order by setting $\mu$ $=$ $2$ $-$ $Ng_c/2$ which 
determines $\gamma_{\Phi,1}(g_c)$ as an explicit function of $g_c$ since 
$\gamma_{\Phi,1}(g_c)$ $=$ $\eta_1/2$. Then it is easy to deduce for 
non-critical couplings that  
\begin{equation}
\gamma_{\Phi}(g) ~=~ \frac{2\Gamma(2-y)}{N\Gamma(2-y/2) \Gamma^2(1-y/2) 
\Gamma(y/2)} ~+~ O\left(\frac{1}{N^2}\right) 
\label{eq:gamexp} 
\end{equation}
where we have introduced the scaled coupling $y$ $=$ $Ng$ for convenience. For
subsequent calculations it is also useful to define a new function $G(y/2)$ by 
\begin{equation} 
G\! \left( \frac{y}{2} \right) ~=~ \frac{\Gamma(2-y)}{\Gamma(2-y/2) 
\Gamma^2(1-y/2) \Gamma(1+y/2)} 
\label{eq:G}
\end{equation} 
Taking into account the symmetry factor of $1/2$ present in each $\Phi$ loop,
(\ref{eq:gamexp}) is precisely the result obtained in \cite{bubble}. We can 
perform a similar exercise for the $\beta$-function. Noting that $\omega_1$ $=$
$2y_c^2 G(y_c/2)$ and defining $\beta(g)$ at $O(1/N)$ as 
\begin{equation}
\beta(g) ~=~ (\mu-2)g ~+~ \frac{1}{2}(N+4)g^2 ~+~ g^2 f_1 \! \left( 
\frac{gN}{2} \right) 
\end{equation}
where $f_1(gN/2)$ is an arbitrary function we see that at the critical point we 
must have
\begin{equation}
0 ~=~ 2(\mu-2) ~+~ (N+4)g_c ~+~ 2g_c f_1 \! \left( \frac{g_cN}{2} \right) 
\label{eq:con}
\end{equation}
Using (\ref{eq:con}) to eliminate the $(N+4)g_c$ term in $\beta^\prime(g_c)$ 
we find that 
\begin{equation} 
f_1^\prime(\epsilon) ~=~ -~ \frac{\omega_1(\epsilon)}{2\epsilon^2}
\end{equation} 
Therefore we find in four dimensions  
\begin{equation}
\beta(g) ~=~ \frac{1}{2}(N+4)g^2 ~-~ 4g^2 \int_0^{gN/2} \! dx \, G(x) ~+~ 
O\left(\frac{1}{N^2}\right)
\label{eq:beta1}
\end{equation}
which was obtained originally in \cite{bubble}. To compare with perturbation
theory one computes the integral by first expanding the integrand in a Taylor 
series as a function of $x$ before integrating each term individually. It is 
elementary and reassuring to check that the coefficients of (\ref{eq:gammaphi})
and (\ref{eq:beta}) are reproduced at four loops respectively from 
(\ref{eq:gamexp}) and (\ref{eq:beta1}). It is worthwhile comparing some of the 
features of both methods that lead to the same results. In \cite{bubble} an 
infinite number of diagrams had to be summed which was only made possible by a 
seemingly miraculous cancellation of terms. By contrast one needs only to 
calculate two one-loop diagrams in the critical approach. The elimination of 
the amplitude factor $z$ between both Dyson equations in effect reproduces the 
conventional bubble sum giving the same results in the end. However, the 
advantage of the critical method lies in performing $O(1/N^2)$ calculations. 
Indeed it is not always clear if the corresponding bubble sums at this order 
would prove to be easily calculable but would certainly rely on further 
remarkable cancellations. We can continue the exercise of converting the 
exponents into the renormalization group functions by considering $\eta_2$. By 
retaining the higher orders in the critical coupling we find that to $O(1/N^2)$
\begin{eqnarray}
\gamma_{\Phi}(g) &=& \frac{y}{N} G \! \left(\frac{y}{2}\right) ~+~ 
\frac{2y}{N^2} \left[ \left( G \! \left(\frac{y}{2}\right) ~+~ \frac{y}{2} 
G^\prime \! \left( \frac{y}{2} \right) \right) \beta_1 \! \left( \frac{y}{2} 
\right) \right.  \nonumber \\ 
&& \left. \quad \quad \quad \quad \quad \quad \quad \quad \quad +~ y G^2 \! 
\left(\frac{y}{2}\right) \left( \Psi(y) - \frac{2}{y} - \frac{1}{(y-1)} 
- \frac{1}{(y-2)} \right) \right]  
\label{eq:ga2}
\end{eqnarray}
where, for later convenience, we have introduced the function $\beta_1(gN/2)$ 
which is defined by 
\begin{equation} 
\beta(g) ~=~ (\mu-2)g ~+~ \frac{Ng^2}{2} ~+~ g^2 \beta_1 \! \left( \frac{gN}{2}
\right) ~+~ O\left( \frac{1}{N^2} \right)  
\label{eq:bet1def} 
\end{equation} 
and represents all the $O(1/N)$ corrections. Its explicit form can be deduced 
from (\ref{eq:beta1}). The function $\Psi(y)$ is given by 
\begin{equation}
\Psi(y) ~=~ \psi(1-y) ~+~ \psi \left( 1 + \frac{y}{2} \right) ~-~ 
\psi \left( 1 - \frac{y}{2} \right) ~-~ \psi(1)
\label{eq:psi}
\end{equation}
and its expansion in powers of $y$ will only involve the Riemann zeta series. 
The higher order correction to (\ref{eq:bet1def}) will be determined later. If 
we expand (\ref{eq:ga2}) to $O(g^4)$ we recover all the coefficients of 
(\ref{eq:gammaphi}) at $O(1/N^2)$ which leads us to believe the result for 
$\eta_2$ is correct. 

We can now address one of the issues raised in sect. 1 concerning the 
equivalence of the Wess-Zumino model with a lower dimensional theory. The 
criterion for determining which model this ought to be is dictated by the 
nature of symmetries and the interaction. At criticality the quadratic terms in
an action primarily determine the canonical dimensions of the fields and 
couplings. Therefore in the case of the $O(N)$ Wess-Zumino model an obvious 
candidate is the $O(N)$ supersymmetric $\sigma$ model which is defined in two
dimensions and has the superspace action  
\begin{equation}
S ~=~ - \, \frac{1}{4} \int d^{d}x \, d^2\theta \, \left[ \frac{1}{2}
\bar{D} \Phi D\Phi ~-~ \sigma \left(\Phi^2 - \frac{1}{\lambda}\right) \right]
\label{eq:sigact} 
\end{equation}
Here the coupling constant $\lambda$ differs from that of (\ref{eq:act}) in 
that it is dimensionless in two dimensions and its critical value is deduced
from its $\beta$-function which will clearly be different from (\ref{eq:beta}).
As this model has also been studied to $O(1/N^2)$ we are in a position to 
compare the critical exponents of the fields in both models. (Comparing the
exponents $\nu$ $=$ $-$ $\beta^\prime(\lambda_c)$ of the supersymmetric
$\sigma$ model with $\omega$ $=$ $-$ $\beta^\prime(g_c)$ of the Wess-Zumino 
model is not sensible as their respective canonical dimensions are $(\mu-1)$ 
and $(\mu-2)$.) From the results of \cite{susysig0,susysig1} it transpires that
the $\Phi$-superfield dimension has the same $d$-dimensional value to 
$O(1/N^2)$. Whilst this was expected at leading order because the diagrams are
the same, the agreement at next order is remarkable when we compare the nature 
of the calculation of $\eta_2$ in both models. In the Wess-Zumino model 
$\eta_2$ is determined purely by iterating the equation which produces $\eta_1$
to the next order because there are no new contributing graphs at $O(1/N^2)$. 
By contrast in the $\sigma$ model to obtain $\eta_2$ one has to compute the 
contribution from two and three loop topologies which are divergent and need to
be regularized in the manner discussed earlier. In the component field 
calculation this equated to evaluating $24$ graphs, \cite{susysig1}. Therefore 
it would appear that both (\ref{eq:act}) and (\ref{eq:sigact}) are equivalent 
at least at $O(1/N^2)$. In other words $\gamma_\Phi(g_c)$ $=$ 
$\gamma_\Phi(\lambda_c)$ to this order. However, the $\sigma$-field of both 
models does not have the same dimension as a direct consequence of the 
non-renormalization theorem. As noted earlier $\chi$ $=$ $0$ in the Wess-Zumino
model but in the $\sigma$ model $\chi$ is non-zero. For example, at leading 
order $\chi_1$ $=$ $-$ $\eta_1$. This property can be illustrated further if we
consider calculating the exponent $\nu$ $=$ $\mu$ $-$ $1$ $+$ $O(1/N)$ in the 
Wess-Zumino model and compare it with the value in  the $\sigma$ model, 
\cite{susysig2}. To do this we can replace the exponent $\omega$ in the 
correction term in each of the scaling forms (\ref{eq:prop}) with $\nu$. 
Consequently the leading order value $\nu_1$ can be deduced from the equation 
analogous to (\ref{eq:omt}) where now both terms on the right side will 
contribute giving  
\begin{equation}
\frac{\eta_1}{(\eta_1 - 2\nu_1)} ~=~ 1 
\end{equation}
which implies $\nu_1$ $=$ $0$. This is the same as the value in the $\sigma$ 
model, \cite{susysig0,susysig1}, and would again have encouraged us to believe
in a possible equivalence. However, unlike the situation with $\omega$ the 
coordinate space version of the matrix consistency equation which determines 
$\nu$ does not have the reordering problem and we can therefore also compute 
$\nu_2$. This is essentially trivial due to the absence of the higher order two
and three loop topologies due to chirality which again means there are no new 
higher order graphs which need to be included. Therefore $\nu_2$ is deduced
by expanding the consistency equation which gave $\nu_1$ to the next order in
$1/N$, in the same way that $\eta_2$ was deduced. We found that $\nu_2$ $=$ $0$.
In the $\sigma$ model $\nu$ has a non-zero contribution at $O(1/N^2)$, 
\cite{susysig2}. It is not clear whether $\nu$ will vanish to all orders in 
$1/N$ in the Wess-Zumino model since there will be new graphs at next order. 
Their contributions would have to cancel to have $\nu_3$ $=$ $0$. In light of 
these remarks it would certainly be very surprising if $\eta_3$ calculated in 
both models was the same since the effect of differing values of $\chi$ would 
almost certainly be important at $O(1/N^3)$. In all these observations 
concerning a possible equivalence the underlying reason for its failure after 
$O(1/N^2)$ is due to the chirality condition which implies $\chi$ $=$ $0$. 
However, another way of understanding it is by considering supersymmetry in 
relation to the space time dimension. Clearly the Wess-Zumino model is 
invariant under ${\cal N}$ $=$ $1$ supersymmetry in four dimensions. However, 
theories in two dimensions which are built out of chiral superfields are 
invariant under ${\cal N}$ $=$ $2$ supersymmetry. Therefore if one wished to 
construct the lower dimensional equivalent theory to (\ref{eq:act}) it would 
need to have more than one supersymmetry. The $O(N)$ supersymmetric $\sigma$ 
model unfortunately does not satisfy this criterion being invariant under 
${\cal N}$ $=$ $1$ supersymmetry which is why it is remarkable that its 
$\eta_2$ and that of the Wess-Zumino model are the same.  

\sect{Calculation of $\omega_2$.}
\label{sec:3}

We are now in a position to extend the result for $\omega_1$ to the next 
order in $1/N$. Unlike the calculation for $\eta_2$ which simply involved the
expansion of (\ref{eq:de}) to next order the reordering which occurs in 
(\ref{eq:om}) means we need to include the contributions from several higher 
order graphs. These are illustrated in fig. 2. In the analogous calculation of
the $\beta$-function in $\phi^4$ theory, \cite{knot,nim}, there were $33$ 
graphs which needed to be computed. In principle the same graphs would have to 
be considered here but the chirality condition excludes the majority of these 
to leave the topologies of fig. 2. Consequently if we denote their total value 
by $\Pi_2$, the solution of the Dyson equation, (\ref{eq:omt}), can be expanded 
to the next order in $1/N$ to obtain 
\begin{equation} 
\omega_2 ~=~ - \left[ 4(\mu - 1)(\mu - 2) \hat{\Psi}(\mu) ~+~ 
\frac{(16\mu^4 - 96\mu^3 + 188\mu^2 - 136\mu + 25)}{2(2\mu - 3)(\mu - 1)} 
\right] \eta_1^2 ~+~ \Pi_2  
\label{eq:om2sd} 
\end{equation} 
where $\hat{\Psi}(\mu)$ $=$ $\psi(2\mu-3)$ $+$ $\psi(3-\mu)$ $-$ $\psi(\mu-1)$ 
$-$ $\psi(1)$. The terms not involving $\Pi_2$ correspond to the expansion of 
the $a$-functions of (\ref{eq:omt}). We note that at $O(1/N^2)$ we can no 
longer neglect the second term of (\ref{eq:om1}). Before we can substitute the 
lines with the critical propagators, we need to compute the $D$-algebra of each 
graph. Elementary application of the rules of \cite{grisa} means that we are 
left with a $\Box$-operator acting on several lines which in momentum space 
will correspond to an extra factor of $p^2$ where $p$ is the momentum flowing 
in that line. We note that in perturbation theory this invariably means that 
that line vanishes from the Feynman integral since massless propagators have 
unit exponent. In the critical large $N$ case this will not always be the case 
as the $\Box$-operator can act on a $\sigma$-line whose canonical exponent is 
not unity. Before detailing the evaluation of each graph we record how the 
$\Box$-operators are distributed around each of the graphs of fig. 2. For the 
non-planar graph each of the upper $\Phi$ lines joining the external $\sigma$ 
lines has a $\Box$-operator, which means in the light of the above remarks that
they now have zero exponent. The analogous lines of the four loop graph also 
have one of these operators acting on them to also reduce them to zero exponent
in momentum space, as does the completely central $\sigma$ line. 

It turns out that for the non-planar graph this $D$-algebra immediately 
reduces it to the simple two loop topology given in fig. 3 where we use the
coordinate space convention of integrating over vertices in the language of
\cite{vas2}. This graph can be reduced to an integral whose value is well 
known if one applies the transformation $\swarrow$ of \cite{vas2}. The result
is an integral originally computed by Chetyrkin and Tkachov, \cite{chet}, 
denoted by $ChT(1,1)$ in the notation of \cite{vas2}. Hence the full 
contribution of the non-planar graph to $\Pi_2$ is  
\begin{equation} 
-~ \frac{3}{2} (\mu-1) \hat{\Theta}(\mu) \eta_1^2 
\label{eq:int1} 
\end{equation}  
where $\hat{\Theta}(\mu)$ $=$ $\psi^\prime(\mu-1)$ $-$ $\psi^\prime(1)$. We 
note that each term of the $\epsilon$-expansion of $\hat{\Theta}(\mu)$, with 
$\mu$ $=$ $2$ $-$ $\epsilon$, is proportional to the Riemann $\zeta$-function 
beginning with $\zeta(3)$. 

The treatment of the remaining graphs is more involved since there are three
$\sigma$-lines where the correction in the asymptotic scaling functions can
be inserted and the distribution of the $\Box$'s around the graph introduce an
asymmetry. Otherwise two of the insertions would be equivalent under a 
symmetry transformation of the integral. Therefore we cannot assume a priori
that their contributing values would be the same. Of the three possibilities
it transpires that that graph where the $\sigma$-line insertion is on the
bottom line of the graph in fig. 2 (with the $\Box$'s on the upper 
$\Phi$-lines) has already been computed in the bosonic $O(N)$ $\sigma$ model 
calculation of its two dimensional $\beta$-function, \cite{vas2}. For 
completeness, we note that the elementary rules of integration by uniqueness, 
\cite{vas2}, after transforming to the momentum representation leaves the three
loop integral illustrated in the first graph of fig. 4. Therefore, we merely 
quote the contribution of this graph is  
\begin{equation} 
-~ \frac{(\mu-1)(2\mu-3)}{2(\mu-2)}  
\left[ \frac{\hat{\Psi}^2(\mu)}{2} ~+~ \frac{\hat{\Phi}(\mu)}{2} ~+~ 
\frac{\hat{\Psi}(\mu)}{(\mu-2)} - 3\hat{\Theta}(\mu) \right] \eta_1^2  
\label{eq:int2} 
\end{equation} 
where $\hat{\Phi}(\mu)$ $=$ $\psi^\prime(2\mu-3)$ $-$ $\psi^\prime(3-\mu)$ $-$
$\psi^\prime(\mu-1)$ $+$ $\psi^\prime(1)$. Like (\ref{eq:int1}) the 
$\epsilon$-expansion of the function of $\mu$ within the brackets also only 
involves the Riemann $\zeta$ series and no rationals. However, here the first 
term involves $\zeta(5)$.  

It turns out in fact that the graph where the $\sigma$-insertion is on the top 
$\sigma$-line relative to the $\Box$'s is equivalent in value to 
(\ref{eq:int2}), at this order in $1/N$. This can be established either by 
explicit evaluation or by a series of conformal manipulations on the three loop
integral which results after the same elementary steps which produced the first
graph of fig. 4. Instead the second graph of fig. 4 is obtained. To relate it 
to the former we first perform a conformal transformation with the origin on 
the right which produces an integral with the same topology but with different
exponents on the lines. Replacing the $(y-u)$ line by a chain integral, 
\cite{vas2}, and choosing the exponents of these two lines in such a way that 
the top right internal vertex is now unique, one obtains a three loop graph 
with the same topology but now with exponents $(\mu-1)$ on all the lines except
for the $(y-u)$ and $(x-z)$ lines whose exponents are $(3-\mu)$. Finally, by 
attaching a propagator of exponent $(2\mu-2)$ to the $x$ external point using 
the chain integration rule of \cite{vas2}, the final integration over this 
unique vertex yields the first graph of fig. 4. Moreover the factors associated
with each of these integration steps reduce to unity. Therefore both these 
insertions are equal in value. 
 
The remaining integral with the insertion on the central $\sigma$-line 
cannot immediately be reduced to one whose computation has been given 
previously. Therefore we will give its evaluation in detail. Transforming to 
the momentum representation yields the first graph of fig. 5. After carrying
out a conformal left transformation on it the resulting integral has a 
unique triangle, so that when it is completed the subsequent integral has a
unique vertex. Performing this and a simple chain integral, before undoing the
original conformal left transformation, one is left with the second integral of
fig. 5. The associated factor is $a^4(\mu-1)/a(2\mu-4)$ relative to the first
graph of fig. 5. Then applying the successive transformations $\rightarrow$
and $\leftarrow$ in the language of \cite{vas2}, before applying the momentum 
representation transformation yields the final graph of fig. 5. Ordinarily in 
perturbation theory when such a topology is encountered where all the lines 
have unit exponent, the way to proceed is to integrate by parts. Although the 
same strategy applies here, it turns out that the set of graphs which result 
from this are individually divergent, though the overall result is clearly 
finite. To handle this one has to introduce a temporary (analytic) regulator 
$\delta$ in the power of various lines. The final result will of course be 
independent of any choice, but it is judicious to choose the distribution of 
$\delta$'s in such a way that subsequent integrations can be carried out. To 
this end we replace the exponents of the lines with $(2\mu-3)$ by 
$(2\mu-3-\delta)$ and the central lines with exponent $1$ by $(1+\delta)$. This
symmetric distribution means that two integrals result after integration by 
parts, each of which can be immediately reduced to a two loop integral where 
there is a common factor proportional to $1/\delta$. After several elementary 
transformations each of these has the $\delta$ $=$ $0$ form $ChT(3-\mu,\mu-1)$.
However, the $1/\delta$ pole means that each $\delta$ $\neq$ $0$ integral needs
to be determined to the $O(\delta)$ term. Unfortunately this is not possible 
for each case. Instead it turns out that due to the distribution of $\delta$'s 
in each integral we can compute the {\em difference} in the values of the 
integrals to the order we require and it is merely given by the value of 
$ChT(3-\mu,\mu-1-\delta)$ to $O(\delta)$. This is achieved by comparing the 
corresponding exponents in each graph. In other words if the lines have the 
same exponent for non-zero $\delta$, then to $O(\delta)$ they will not 
influence the value of the difference. Therefore in these lines we can 
effectively set $\delta$ $=$ $0$. Only for those lines where there is a 
mismatch in the exponent in corresponding lines must one not set $\delta$ $=$ 
$0$ in them. After matching in this way, one essentially reduces the 
calculation to the evaluation of $[ChT(3-\mu,\mu-1-\delta)$ $-$ 
$ChT(3-\mu,\mu-1)]/\delta$ as $\delta$ $\rightarrow$ $0$. Thus piecing the 
steps in this summary together, the total contribution to $\Pi_2$ from this 
integral is, including the amplitude factors, 
\begin{equation} 
-~ \frac{(\mu-1)(2\mu-3)}{2(\mu-2)} \left[ \hat{\Phi}(\mu) ~+~ 
\hat{\Psi}^2(\mu) ~-~ \frac{8(2\mu-3)}{(\mu-1)(\mu-2)\eta_1} ~-~ 
\frac{2}{(\mu-2)^2} \right] \eta_1^2  
\label{eq:int3} 
\end{equation} 
Although the function within the brackets is different from (\ref{eq:int2}) its
$\epsilon$-expansion is similar in that it begins with $\zeta(5)$.   

Finally, it is an elementary exercise now to substitute the values for these
integrals into (\ref{eq:om2sd}) to find 
\begin{eqnarray} 
\omega_2 &=& \eta^2_1 \left[ \frac{3(3\mu - 4)(\mu - 1)\hat{\Theta}(\mu)}
{2(\mu-2)} ~-~ \frac{(2\mu - 3)(\mu - 1) \hat{\Psi}^2(\mu)}{(\mu - 2)} 
{}~+~ \frac{4(2\mu - 3)^2}{(\mu - 2)^2 \eta_1} \right. \nonumber \\ 
&& -~ \left. \frac{(4\mu^3 - 24\mu^2 + 50\mu - 35)(\mu-1) \hat{\Psi}(\mu)} 
{(\mu-2)^2} ~-~ \frac{(2\mu - 3)(\mu - 1) \hat{\Phi}(\mu)}{(\mu - 2)} 
\right. \nonumber \\ 
&& +~ \left. \frac{(16\mu^7 - 192\mu^6 + 956\mu^5 - 2536\mu^4 + 3825\mu^3 
- 3212\mu^2 + 1328\mu - 182)}{2(\mu - 1)(\mu - 2)^3(2\mu-3)} \right] 
\label{eq:omexp} 
\end{eqnarray} 
We draw attention to the fact that as in other models where there is a 
reordering of the Dyson equation to produce the $\beta$-function, \cite{gn2},
there is a term linear in $\eta_1$ which prevents the $O(1/N^2)$ correction 
being proportional to $\eta^2_1$. 

\sect{Results and discussion.} 

Following the same procedure we used to determine $\gamma_\Phi(g)$ to 
$O(1/N^2)$, we can write down the $\beta$-function in a similar way. If we 
define the $O(1/N^2)$ $\beta$-function by 
\begin{equation} 
\beta(g) ~=~ (\mu-2)g ~+~ \frac{Ng^2}{2} ~+~ g^2 \beta_1 \! \left( \frac{gN}{2}
\right) ~+~ \frac{g^2}{N} \beta_2 \! \left( \frac{gN}{2} \right) ~+~ 
O\left( \frac{1}{N^3} \right)  
\end{equation} 
then a similar derivation to that which produced (\ref{eq:beta1}) gives 
\begin{equation} 
\beta_2 \! \left( \frac{gN}{2} \right) ~=~ 2 \beta_1^2 \! \left( \frac{gN}{2} 
\right) ~-~ 8 ~+~ \frac{1}{2} \int_0^{gN} dy \left[ y \beta_1 \! 
\left( \frac{y}{2} \right) \beta_1^{\prime\prime} \! \left( \frac{y}{2} 
\right) ~-~ \frac{2\Omega_2(y)}{y^2} \right] 
\label{eq:beta2} 
\end{equation} 
where  
\begin{eqnarray}
\Omega_2(y) &=& 4 \left[ (y-2)(y-1)y \left( \!\Psi^2(y) + \Phi(y) 
- \left( \frac{2}{y} + \frac{y^2}{(y-1)}\right) \! \Psi(y) 
- \frac{3}{4} \left( 3 - \frac{1}{(y-1)} \right) \Theta(y) \! \right) \right. 
\nonumber \\ 
&& \left. ~~~~-~ \frac{y(4y^2-7y+4)}{(y-2)} - \frac{y^2}{(y-1)} ~-~ \frac{8}{y}
{}+ (y-2)(y^3+y^2-6) \right] G^2 \! \left(\frac{y}{2}\right) \nonumber \\  
&& +~ \frac{32(y-1)^2}{y} G\! \left(\frac{y}{2} \right) 
\label{eq:omy} 
\end{eqnarray}
and the new functions $\Phi(y)$ and $\Theta(y)$ are given by
\begin{eqnarray}
\Phi(y) &=& \psi^\prime(1-y) ~-~ \psi^\prime \left( 1 + \frac{y}{2} \right) ~-~ 
\psi^\prime \left( 1 - \frac{y}{2} \right) ~+~ \psi^\prime(1) \nonumber \\
\Theta(y) &=& \psi^\prime \left( 1 - \frac{y}{2} \right) ~-~ \psi^\prime(1) 
\end{eqnarray}
After expanding the integrand in powers of $y$ and completing the integral, it
is easy to verify that (\ref{eq:omy}) reproduces the first four coefficients of
(\ref{eq:beta}) at the appropriate order in $1/N$ which gives us confidence
that (\ref{eq:omexp}) is correct. Having therefore verified that our result is
consistent with all known information we can now provide some new 
coefficients in the perturbative series in four dimensions beyond 
(\ref{eq:beta}). For instance, if we represent the unknown $O(1/N^2)$ higher
order contributions to the $\beta$-function by  
\begin{equation} 
\beta(g) ~=~ \frac{1}{2}(d-4)g ~+~ \frac{1}{2}(N+4)g^2 ~+~ \sum_{r=2}^\infty 
(a_rN+b_r)N^{r-2}g^{r+1} 
\end{equation}  
then we can deduce 
\begin{eqnarray} 
a_5 &=& \frac{1}{16} ~-~ \frac{\zeta(3)}{8} \nonumber \\
a_6 &=& \frac{1}{40} ~+~ \frac{\zeta(3)}{20} ~-~ \frac{3\zeta(4)}{40} 
\nonumber \\ 
b_5 &=& \frac{25}{24} ~+~ \left( \frac{13}{8} ~-~ \frac{9}{4}\zeta(3) 
\right) \zeta(3) ~-~ \frac{27}{16}\zeta(4) ~+~ \frac{39}{2}\zeta(5) ~-~
\frac{75}{8}\zeta(6) \nonumber \\
b_6 &=& \frac{33}{40} ~-~ \left( \frac{11}{20} ~-~ \frac{51}{10}\zeta(3) ~+~
\frac{27}{10}\zeta(4) \right)\zeta(3) ~+~ \frac{27}{40}\zeta(4) \nonumber \\
&& -~ \frac{57}{5}\zeta(5) ~-~ \frac{153}{8}\zeta(6) ~-~ 11\zeta(7) 
\end{eqnarray} 
The five loop coefficients, which are new, will provide a useful crosscheck on
future full perturbative calculations in $\MSbar$ in this model. 

Ordinarily once an exponent has been checked for consistency with perturbation
theory one can gain additional information about the same exponent in the three 
dimensional theory since the expression is usually valid in $d$-dimensions. 
However, attempting this for (\ref{eq:omexp}) one finds that $\omega_2$ 
diverges as $\mu$ $\rightarrow$ $3/2$. This is due to non-cancelling 
singularities arising from the functions $\hat{\Phi}(\mu)$, $\hat{\Psi}^2(\mu)$ 
and the final term. If one compares this to the situation in $O(N)$ $\phi^4$ 
theory this is somewhat unexpected since its $\omega_2$ is well behaved in 
three dimensions, \cite{knot,nim}. For (\ref{eq:omexp}) one can trace the 
source of the terms which are divergent in three dimensions to the original 
Feynman integrals. It turns out that they arise solely from the four loop 
integral of fig. 2. In the calculation of \cite{knot} there is an additional 
singularity from the graph which is a non-planar version of the four loop graph
of fig. 2. Its contribution in $O(N)$ $\phi^4$ theory precisely cancels the 
singularity from the planar four loop graph. As we have already noted the 
non-planar topology does not arise in the Wess-Zumino model due to chirality 
and it is its absence here which leads to a singular $\omega_2$. In making this
comparison it should be pointed out that whilst the explicit $\mu$-dependent 
values of the analogous four loop integral of fig. 2 will in general be 
different in both models, each will have the same combination of divergent
functions, $\hat{\Phi}(\mu)$ and $\hat{\Psi}^2(\mu)$, with the same residues as
$\mu$ $\rightarrow$ $3/2$.  

In producing expressions for the $\beta$-function like (\ref{eq:beta1}) and 
(\ref{eq:beta2}) we have so far assumed that they are always evaluated by first 
expanding the integrand in a power series about the lower end of the 
integration range. Each term of the series is then integrated separately. 
However, we can also regard these expressions as being a summation of part of
the full (unknown) perturbative series for $\beta(g)$. Although the explicit 
integration cannot be performed even for $\beta_1(gN/2)$ we can at least 
determine some information on the range of validity of these expressions and
how the subsequent corrections affect it. As was pointed out in \cite{bubble}
the obstruction to the resummation at $O(1/N)$ is given by the first 
singularity in the integrand of $\beta_1(gN/2)$. This was found to be at $g$ 
$=$ $3/N$ due to the singularity in the numerator $\Gamma$-function of 
$G(y/2)$, (\ref{eq:G}). Although this is also divergent at $y$ $=$ $2$ there is
a compensating pole in the denominator of $G(y/2)$ at the same point. In other
words,  
\begin{equation}
G(y) ~  
\begin{array}{c} 
\\ \sim \\ 
y \rightarrow 3/2 
\end{array} ~ \frac{1}{3\pi^2}\frac{1}{(2y-3)}
\label{eq:pole1} 
\end{equation}
Therefore to $O(1/N)$ $\beta(g)$ diverges logarithmically to $+$ $\infty$ as 
$g$ $\rightarrow$ $3/N$. Moreover, it is also positive in this interval.  
Repeating this analysis now when the explicit expression for $\beta_2(gN/2)$ is
taken into account we observe that the first singularity is now at $g$ $=$ 
$1/N$  It arises purely from the $\Psi^2(y)$ and $\Phi(y)$ terms of 
$\Omega_2(y)$ and in particular 
\begin{equation}
\Omega_2(y) ~  
\begin{array}{c} 
\\
\sim \\ 
y \rightarrow 1
\end{array} ~-~ \frac{128}{\pi^4}\frac{1}{(y-1)}
\label{eq:pole2} 
\end{equation}
which again leads to a logarithmic singularity in $\beta(g)$ but now at $g$ $=$
$1/N$ and tending to negative infinity. Therefore the effect of including the 
higher order corrections is that the obstruction to the resummation moves 
closer to the origin. Moreover, this singularity is also related to the absence
of a finite value for $\omega_2$ in the three dimensional model. This is simply
because the leading order relation between the coupling and $d$ is $\mu$ $=$ 
$2$ $-$ $y/2$. It is also interesting to compare these results with the 
analogous ones for the $O(N)$ $\phi^4$ $\beta$-function at $O(1/N^2)$. From the
expressions for $\omega_1$ and $\omega_2$ given in \cite{knot,nim} and the 
perturbative $\beta$-function of \cite{vlad}, we find that $\beta_1$ has a pole
for $y$ $=$ $15$, whereas $\beta_2$ is singular for $y$ $=$ $9$. Again, the 
pole in $\beta_2$ stems from the (analogous) $[\Psi^2(y)$ $+$ $\Phi(y)]$ term. 
The reason why the coupling range is so much bigger in the $\phi^4$ model is 
simply that the coefficients of the one-loop terms (which determine the 
arguments of the functions $G$, $\Psi$, $\Phi$ and $\Theta$) are $1/2$ and 
$1/6$ in the Wess-Zumino and $\phi^4$ $\beta$-functions respectively. Moreover
the signs of corresponding integrand poles are the same as (\ref{eq:pole1}) and
(\ref{eq:pole2}). Therefore in both models the radius of convergence of the
large $N$ $\beta$-function decreases when higher order corrections are 
included. Further, at $O(1/N^2)$ a spurious fixed point emerges since both 
models are non-asymptotically free and hence their $\beta$-functions are 
positive for very small couplings. In explicit perturbative calculations 
similar zeros can also occur when higher order terms are successively added. 
Since their appearance fluctuates with the number of terms included in the 
series we would expect that the zero at $O(1/N^2)$ would disappear if the 
$O(1/N^3)$ correction was included. In this case it also would be interesting 
to ascertain where the first integrand pole occurs. 

We conclude with several comments. Although as we have discussed there is not a 
full equivalence between the $O(N)$ Wess-Zumino model and the $O(N)$ 
supersymmetric $\sigma$ model we can make some observations on the numerology 
of the wave function exponent in both models and examine the effect 
supersymmetry has on $\eta_2$. By this we mean the presence of rational and 
irrational coefficients that appear in the corresponding renormalization group 
function. For example, the $\beta$-function of the Wess-Zumino model clearly 
contains rationals and the Riemann zeta series. Likewise if one expands 
$\eta_2$ to higher order rational numbers will occur. Although this may not 
appear surprising, the $\epsilon$-expansion of $\eta_2$ for the $O(N)$ 
supersymmetric $\sigma$ model near $d$ $=$ $2$ $-$ $\epsilon$ dimensions, which
is the same as the result for the Wess-Zumino model, yields only the series 
$\zeta(n)$, $n$ $\geq$ $3$, beyond two loops to all orders in perturbation 
theory. The $\beta$-function at $O(1/N^2)$ has the same property, 
\cite{susysig2}. By contrast the bosonic $O(N)$ $\sigma$ model has rational 
coefficients at two and higher loops in its renormalization group functions, 
\cite{sig,alv}, and the imposition of supersymmetry renders the cancellation of
the rationals by fermionic partner graphs. Naively one might expect a similar 
property for four dimensions. However, in rewriting the $\psi(x)$ functions of 
(\ref{eq:eta2}) to ensure their $\epsilon$-expansion has only $\zeta(n)$ terms 
now with respect to $d$ $=$ $4$ $-$ $2\epsilon$, one introduces functions whose
expansion will produce rationals. This arises from the iteration of the Hartree
Fock approximation and, at least for the Wess-Zumino $\beta$-function, we have 
shown at $O(1/N^2)$ that the rational coefficients in the four dimensional 
result also arise from a similar iteration. The higher order graphs included 
here involve only $\zeta(n)$'s in their expansion as we emphasised in their 
computation. Indeed in light of these remarks several questions arise as to 
whether the imposition of another supersymmetry on this model would reduce the 
renormalization group functions of that model to rational at one loop only and 
irrational thereafter. Also, it is natural to question whether {\em all} the 
higher order graphs which remain in the Wess-Zumino model after the chirality 
condition has been imposed and the $D$-algebra performed yield integrals whose 
$\epsilon$-expansion involve irrationals or non-zeta transcendentals as occurs 
in the $O(N)$ $\phi^4$ theory, \cite{knotbk,knot}. One area where such issues 
could be addressed further is in the calculation of higher order corrections to
some of the other critical exponents. We would hope to return to this issue 
later. 
 
\vspace{0.25cm} 
{\bf Acknowledgements.} This work was carried out with support from PPARC
through an Advanced Fellowship, (JAG), and JNICT by a scholarship, (PF). We
also thank Prof D.R.T. Jones and Dr I. Jack for several useful discussions. 

\appendix

\newpage 
{\epsfxsize=12cm 
\epsfbox{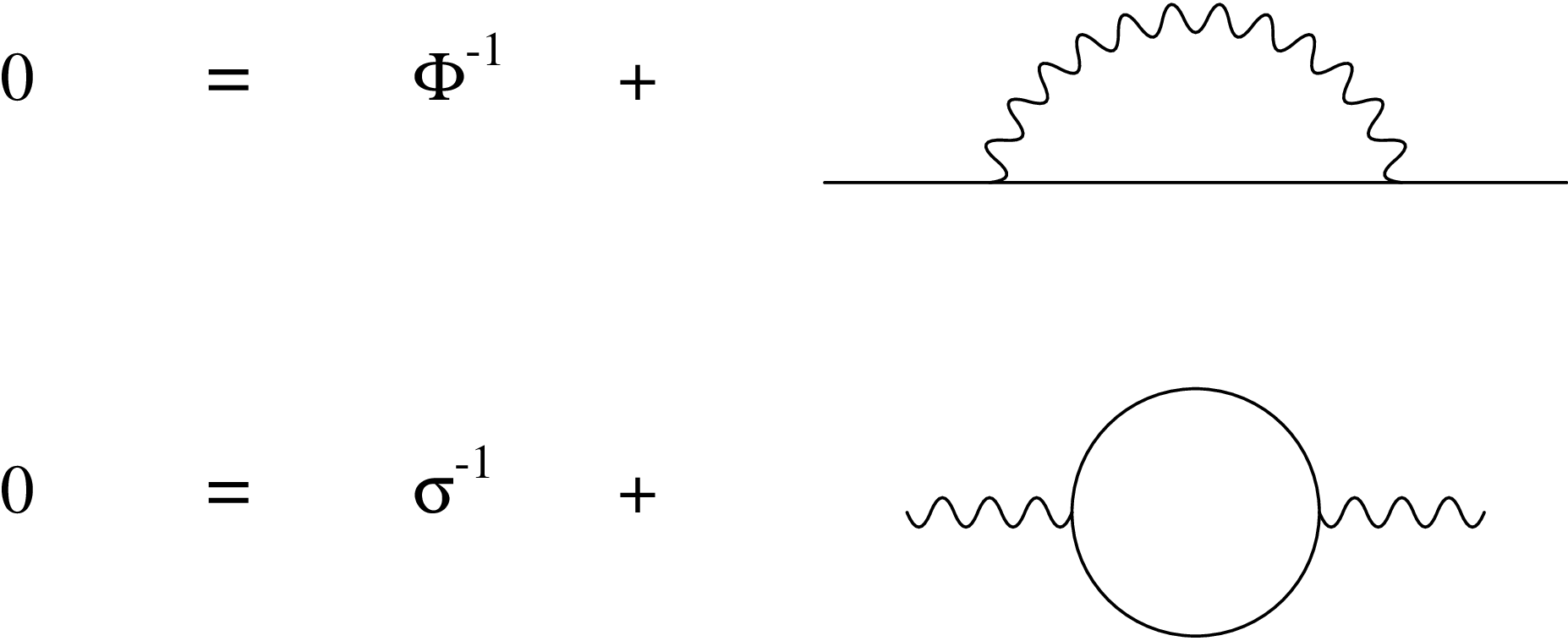}}  
\vspace{1cm} 
{\bf Fig. 1. Leading order Schwinger Dyson equations.} 

\vspace{3cm} 
{\epsfysize=3cm 
\epsfbox{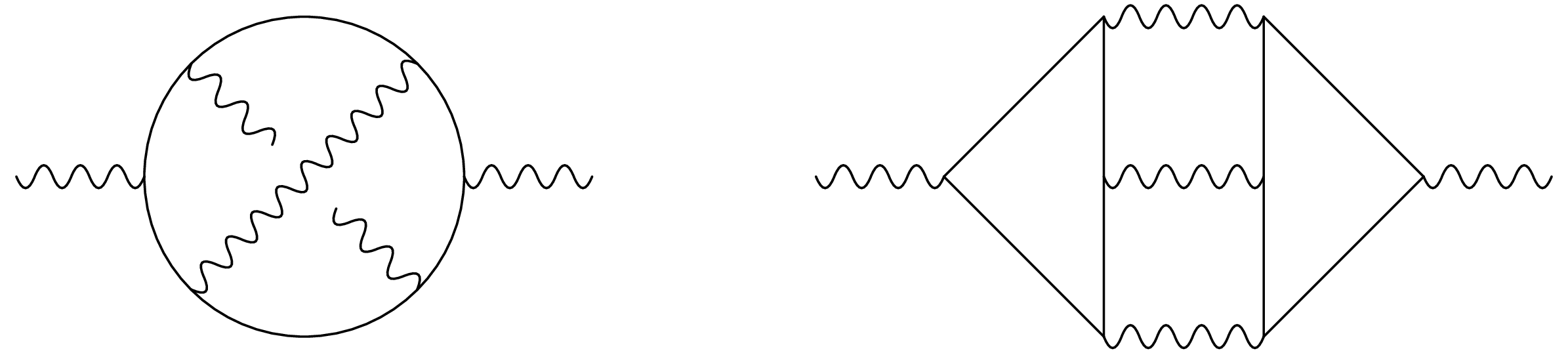}}  
\vspace{1cm} 
{\bf Fig. 2. Additional graphs for the $\sigma$ Dyson equation.} 

\vspace{3cm} 
{\epsfysize=3cm 
\epsfbox{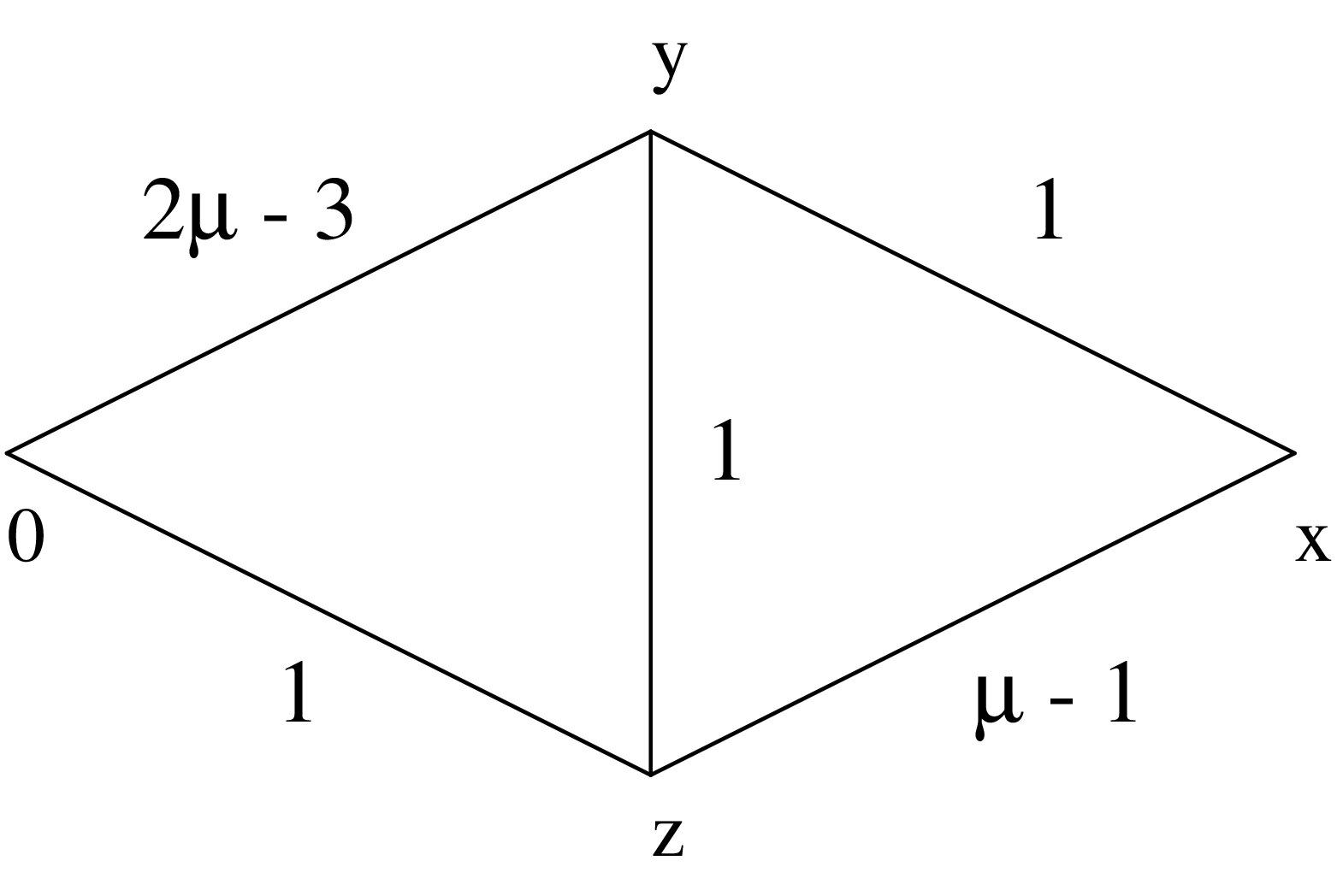}}  
\vspace{1cm} 
{\bf Fig. 3. Intermediate integral in the calculation of the non-planar graph.} 

\newpage 
{\epsfysize=5cm 
\epsfbox{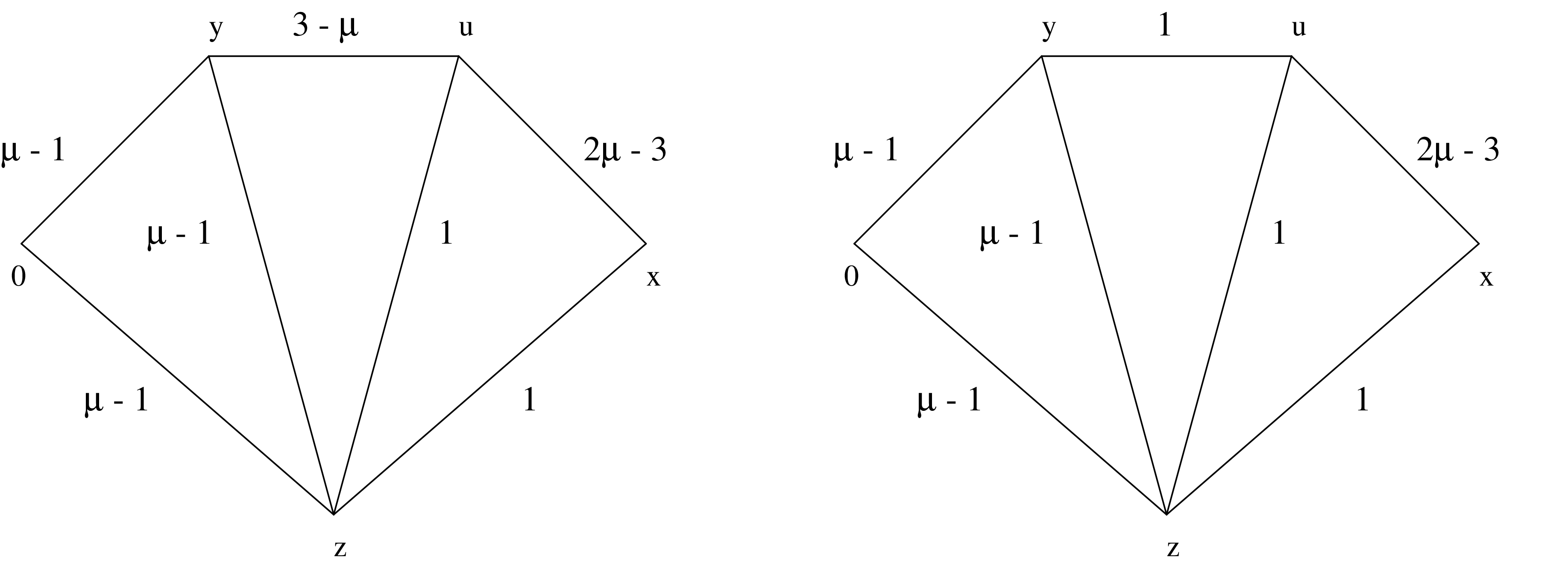}}  
\vspace{1cm} 
{\bf Fig. 4. Equivalent three loop integrals.} 

\vspace{3cm} 
{\epsfysize=10cm 
\epsfbox{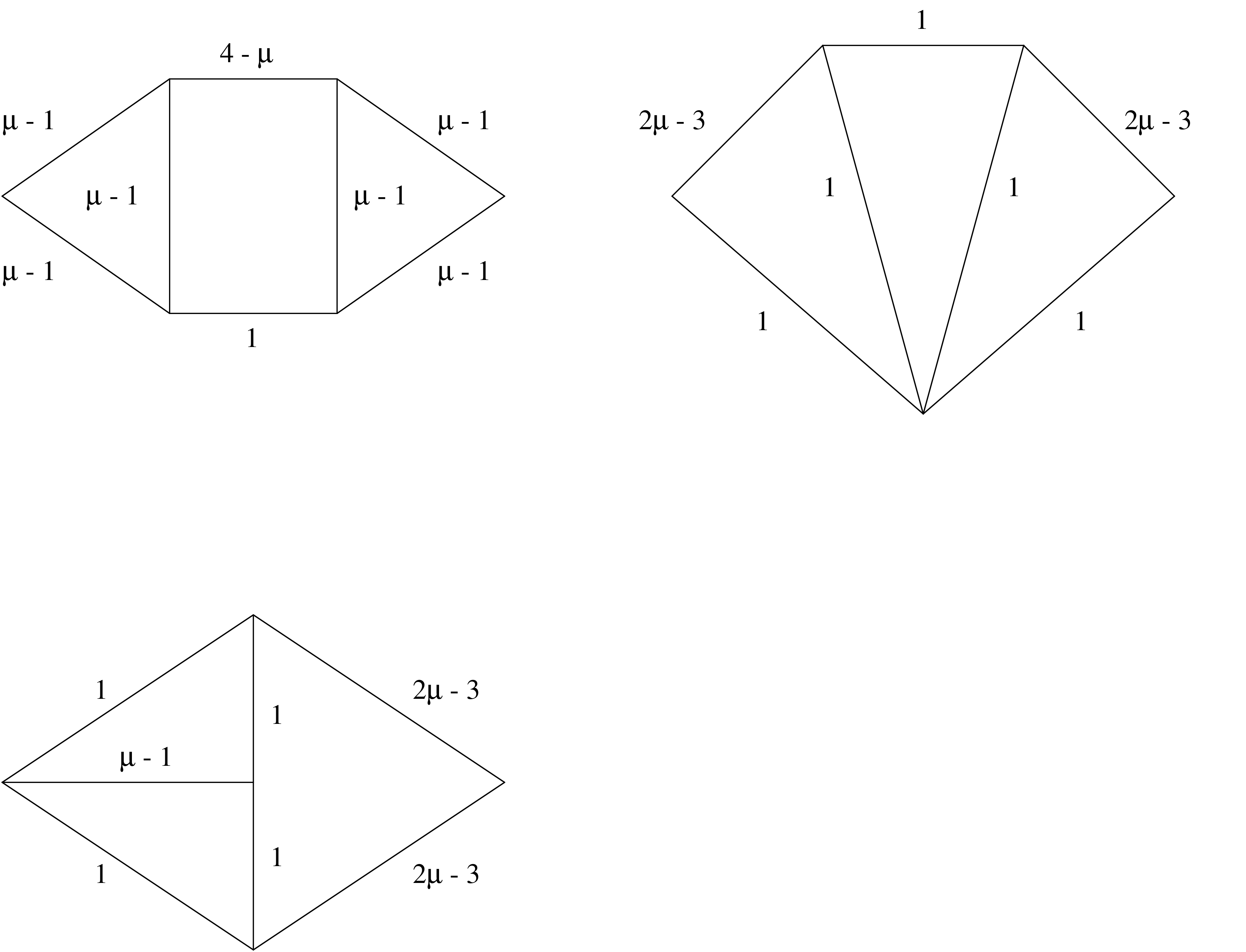}}  
\vspace{1cm} 
{\bf Fig. 5. Intermediate integrals in the calculation of the four graph with 
insertion on the central line.} 


\newpage 
\begin{thebibliography}{99}
\bibitem{wz} J. Wess \& B. Zumino, Phys. Lett. {\bf 49B} (1974), 52.  
\bibitem{alv} L. Alvarez-Gaum\'{e}, D.Z. Freedman \& S. Mukhi, Ann. Phys. 
{\bf 134} (1981), 85.  
\bibitem{sig} D. Friedan, Phys. Rev. Lett. {\bf 45} (1980), 1057.  
\bibitem{nonren} J. Iliopoulos \& B. Zumino, Nucl. Phys. {\bf B76} (1974), 310;
S. Ferrara, J. Iliopoulos \& B. Zumino, Nucl. Phys. {\bf B77} (1974), 413.  
\bibitem{earlyloop} P.K. Townsend \& P. van Nieuwenhuizen, Phys. Rev. {\bf D20}
(1979), 1832; A. Sen \& M.K. Sundaresan, Phys. Lett. {\bf B101} (1981), 61.  
\bibitem{grisa} L.F. Abbott \& M.T. Grisaru, Nucl. Phys. {\bf B169} (1980), 
415. 
\bibitem{avdeev} L.V. Avdeev, S.G. Gorishny, A.Yu. Kamenshchick \& S.A. Larin, 
Phys. Lett. {\bf B117} (1982), 321.
\bibitem{wzssm} P.M. Ferreira, I. Jack \& D.R.T. Jones, Phys. Lett. {\bf B392} 
(1997), 376.
\bibitem{bubble} P.M. Ferreira, I. Jack \& D.R.T. Jones, Phys. Lett. {\bf B399}
(1997), 258.
\bibitem{vas1} A.N. Vasil'ev, Yu.M. Pis'mak \& J.R. Honkonen, Theor. Math. 
Phys. {\bf 46} (1981), 157. 
\bibitem{vas2} A.N. Vasil'ev, Yu.M. Pis'mak \& J.R. Honkonen, Theor. Math. 
Phys. {\bf 47} (1981) 291.
\bibitem{susysig1} J.A. Gracey, Nucl. Phys. {\bf B348} (1991), 714. 
\bibitem{susysig2} J.A. Gracey, Nucl. Phys. {\bf B352} (1991), 183.
\bibitem{susycpn} M. Ciuchini \& J.A. Gracey, Nucl. Phys. {\bf B454} (1995),
103. 
\bibitem{zinn} J. Zinn-Justin, ``Quantum field theory and critical 
phenomena'' (Clarendon Press, Oxford, 1989).  
\bibitem{susysig0} J.A. Gracey, J. Phys. {\bf A23} (1990), 2183.  
\bibitem{par} M. d'Eramo, L. Peliti \& G. Parisi, Lett. Nuovo Cim. {\bf 2} 
(1971), 878.  
\bibitem{qgraph} P. Nogueira, J. Comput. Phys. {\bf 105} (1993), 279. 
\bibitem{super} S.J. Gates Jr., M.T. Grisaru, M. Ro\v{c}ek \& W. Siegel, 
``Superspace or one thousand and one lessons in supersymmetry'' Frontiers in 
Physics (Benjamin Cummings, Reading, 1983).  
\bibitem{gn2} A.N. Vasil'ev \& A.S. Stepanenko, Theor. Math. Phys. {\bf 97} 
(1993), 1349; J.A. Gracey, Int. J. Mod. Phys. {\bf A9} (1994), 727.  
\bibitem{knot} D.J. Broadhurst, J.A. Gracey \& D. Kreimer, Z. Phys. {\bf C75} 
(1997), 559. 
\bibitem{nim} J.A. Gracey, Nucl. Inst. Meth. {\bf A389} (1997), 361.
\bibitem{chet} K.G. Chetyrkin \& F.V. Tkachov, Nucl. Phys. {\bf B192} (1981),
159. 
\bibitem{vlad} A.A. Vladimirov, D.I. Kazakov \& O.V. Tarasov, Sov. Phys. JETP 
{\bf 50} (1979), 521.
\bibitem{knotbk} D.J. Broadhurst \& D. Kreimer, Int. J. Mod. Phys. {\bf C6} 
(1995), 519.  
\end{thebibliography}
\end{document}